\documentclass[pra,aps,twocolumn,superscriptaddress,floatfix,citeautoscript,showpacs]{revtex4}
\usepackage{graphicx,rotating,subfigure,amsmath,amsfonts,amssymb,delarray}
\usepackage{dsfont}
\usepackage[T1]{fontenc}
\renewcommand{\vec}[1]{\boldsymbol #1}
\newcommand{\e}{\text{e}}

\newcommand{\be}{\begin{equation}}
\newcommand{\ee}{\end{equation}}
\newcommand{\bea}{\begin{eqnarray}}
\newcommand{\eea}{\end{eqnarray}}

\newcommand{\rmd}{d}
\newcommand{\Tr}{Tr}
\allowdisplaybreaks

\begin{document}
\bibliographystyle{apsrev}
\title{Locality and thermalization in closed quantum systems}

\author{J. Sirker}
\affiliation{Department of Physics and Research Center OPTIMAS,
  Technical University Kaiserslautern,
  D-67663 Kaiserslautern, Germany}
\affiliation{Department of Physics and
  Astronomy, University of Manitoba, Winnipeg, Canada R3T 2N2}
\author{N. P. Konstantinidis}
\affiliation{Department of Physics and Research Center OPTIMAS,
  Technical University Kaiserslautern,
   D-67663 Kaiserslautern, Germany}
\author{F. Andraschko}
\affiliation{Department of Physics and Research Center OPTIMAS,
  Technical University Kaiserslautern,
   D-67663 Kaiserslautern, Germany}
\affiliation{Department of Physics and
  Astronomy, University of Manitoba, Winnipeg, Canada R3T 2N2}
\author{N. Sedlmayr\footnote{Current address: Institute de Physique Th\'eorique, CEA/Saclay,
Orme des Merisiers, 91190 Gif-sur-Yvette Cedex, France}}
\affiliation{Department of Physics and Research Center OPTIMAS,
  Technical University Kaiserslautern,
   D-67663 Kaiserslautern, Germany}
\date{\today}

\begin{abstract}
  We derive a necessary and sufficient condition for the
  thermalization of a local observable in a closed quantum system
  which offers an alternative explanation, independent of the
  eigenstate thermalization hypothesis, for the thermalization
  process. We also show that this approach is useful to investigate
  thermalization based on a finite-size scaling of numerical data. The
  condition follows from an exact representation of the observable as
  a sum of a projection onto the local conserved charges of the system
  and a projection onto the non-local ones. We show that
  thermalization requires that the time average of the latter part
  vanishes in the thermodynamic limit while time and statistical
  averages for the first part are identical. As an example, we use
  this thermalization condition to analyze exact diagonalization data
  for a one-dimensional spin model. We find that local correlators do
  thermalize in the thermodynamic limit although we find no
  indications that the eigenstate thermalization hypothesis applies.
\end{abstract}


\maketitle

\section{Introduction}

Preparing a generic classical many-body system in a {\it typical}
initial configuration and letting it time evolve usually leads in the
thermodynamic limit (TDL), to an equilibration at long times so that
{\it typical} observables become time independent. If the dynamics is
ergodic, the {\it ergodic theorem}
\cite{Birkhoff,Neumann1932,Landau1980,MorandiNapoli} ensures that the
time average of observables can be replaced by an ensemble average.
The ensemble provides a probability measure $\rho$ on phase-space
which has to be invariant under time evolution because it describes
the stationary state. $\rho$ therefore has to be a function of the
conserved quantities $\mathcal{Q}_j$ with
$\{\mathcal{H},\mathcal{Q}_j\}=0$, where $\{.,.\}$ is the Poisson
bracket and $\mathcal{H}$ the Hamilton function of the system. In
cases where $\mathcal{H}$ is the only independent conserved quantity
this invariant phase-space measure is the familiar microcanonical
ensemble which becomes equivalent to the canonical one in the TDL.
Integrable systems with phase-space dimension $2N$, on the other hand,
have by definition $N$ independent conservation laws making them
non-ergodic and restricting the motion in phase-space to invariant
tori.
Essential for our understanding of thermalization in classical systems
is the Kolmogorov-Arnold-Moser (KAM) theorem which describes the
consequences of small integrability breaking perturbations on a
quantitative level \cite{MorandiNapoli,Tabor1989}.

Although recent theoretical
\cite{Deutsch1991,Srednicki1994,Rigol2008,Calabrese2011,SirkerGebhard,Sedlmayr2013a}
and experimental
\cite{KinoshitaWenger,TrotzkyChen,HofferberthLesanovsky,StrohmaierGreif}
studies have led to new interesting insights, no equivalent of the
KAM theorem or even a general theory, how and under which conditions
thermalization occurs in the quantum case, exist. A particularly
active field of research has been the investigation of quenches for
lattice models with short-range interactions
\cite{Rigol2008,Calabrese2011,Cazalilla,Rigol2007,EnssSirker,KollathLauchli,RigolMuramatsu,BiroliKollath,ManmanaWessel,Canovi2011,SantosPolkovnikov}.
In this case the initial energy distribution will be {\it
  singly-peaked} in the TDL with vanishing width \cite{Rigol2008}
which is an essential prerequisite to allow for thermalization. A
possible explanation for the thermalization process following a quench
is the {\it eigenstate thermalization hypothesis} (ETH)
\cite{Deutsch1991,Srednicki1994,Rigol2008,Ikeda2011,Rigol2012} which
assumes that the expectation values of an observable in the eigenstate
basis of the Hamiltonian under which the system time evolves fluctuate
little between eigenstates close in energy and can thus be directly
replaced by an ensemble average.

While fulfilling the ETH is sufficient for thermalization for a
generic quench case \cite{note1}
  it is not
a necessary condition. In this article we derive a necessary and
sufficient condition by projecting the considered local observable
onto a part protected by local conservation laws and an orthogonal
part. The condition we derive shows that thermalization can, in
principle, also occur in a more generic scenario when fluctuations
between eigenstate expectation values close in energy are large.

\begin{figure}
\begin{center}
\includegraphics*[width=0.95\columnwidth]{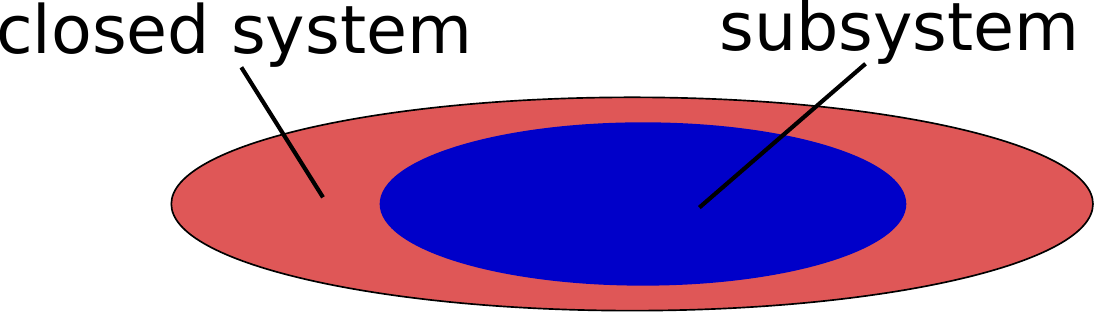}
\end{center}
\caption{We want to consider here local observables acting in a
  subsystem of an infinitely large closed quantum system
  (thermodynamic limit). While the wave function of the full system
  $|\Psi(t)\rangle$ remains a pure state, the reduced density matrix
  $\rho_{\rm red}(t)$ for the subsystem can become thermal for
  $t\to\infty$, implying that all local observables in the subsystem
  thermalize.}
\label{Schematics}
\end{figure}

We want to consider a closed quantum system prepared in an initial
pure state $|\Psi_0\rangle$ time-evolving under a time-independent
Hamiltonian $H$. Clearly the closed quantum system as a whole can
never thermalize because $|\Psi(t)\rangle =\exp(-iHt)|\Psi_0\rangle$
always remains a pure state. The question one can ask, though, and which
we consider here is whether expectation values $\langle O(t)\rangle
=\langle\Psi(t)|O|\Psi(t)\rangle$ of {\it local} observables $O$
acting in a subsystem of an infinitely large closed quantum system,
see Fig.~\ref{Schematics}, will equilibrate at long times with the
equilibrium value $\langle O(t\to\infty)\rangle$ being equal to the
expectation value with respect to the appropriate statistical
ensemble.
More specifically, we will restrict our discussion to time-independent
Hamiltonians $H$ with short-range interactions in position space \cite{note2}.

We first consider the case of a finite dimensional Hilbert space,
e.g., a Hamiltonian acting on a finite lattice. 
The time average of an observable $O$ for an initial normalized pure
state $|\Psi_0\rangle$ is defined by
\begin{equation}
\label{time_ave}
\overline{O}\equiv\lim_{\tau\to\infty}\frac{1}{\tau}\int_0^\tau \rmd t\; \langle\Psi_0|\e^{iHt} O\e^{-iHt}|\Psi_0\rangle.
\end{equation}
By using a spectral representation  of the observable $O$ and assuming
that  the spectrum of $H$ is non-degenerate \cite{note3}
we immediately obtain
\begin{equation}
\label{diag}
\overline{O}=\langle \Psi_0|O_{\rm{diag}}|\Psi_0\rangle = \sum_n O_{nn} \underbrace{|\langle\Psi_0|n\rangle|^2}_{\equiv |c(n)|^2}
\end{equation}
where $O_{\rm{diag}}=\sum_n O_{nn} P_n$.  Here $O_{nn}=\langle
n|O|n\rangle$ with $P_n=|n\rangle\langle n|$ being the projection
operator onto the eigenstate $|n\rangle$ and $H|n\rangle
=\varepsilon_n|n\rangle$. Off-diagonal elements of $O$ do not
contribute in statistical or time averages so that it is sufficient in
the following to consider only the diagonal part $O_{\rm diag}$. To
study thermalization, we are interested in the limit
$\lim_{t\to\infty}\lim_{N\to\infty} \langle O(t)\rangle$, i.e., taking
the TDL, $N\to\infty$, first and only then the limit $t\to\infty$. An
equilibration is only possible in the TDL while $\langle O(t)\rangle$
will show revivals and recurrences in any finite system. An important
question then is if
\begin{equation}
\label{limits}
\lim_{t\to\infty}\lim_{N\to\infty} \langle O(t)\rangle = \lim_{N\to\infty}\overline{O}
\end{equation}
holds true, i.e., if the equilibrium value can be obtained by first
averaging over time and then taking the TDL. In general this is
indeed not the case and explicit counterexamples for bilinear
Hamiltonians such as the one-dimensional transverse Ising model are
known \cite{VenutiZanardiPRA10}. However, for large {\it interacting}
systems one expects that $\langle O(t)\rangle$ is close to
$\overline{O}$ for almost all times, i.e., that the variance is small
\cite{Reimann08}.  In this case Eq.~\eqref{limits} is expected to
hold. We assume in the following that this is the case as is also
tacitly assumed in most other recent thermalization studies.

The attempt to replace the time average by an ensemble average
following the prescription in the classical case immediately leads to
a crucial aspect which is different from the classical problem: All
the projection operators are conserved, $[H,P_n]=0$, and the number of
conserved quantities thus always equals the Hilbert space dimension
$D$. A density matrix $\rho_{\rm{D}}$ which yields a statistical
average equal to the time average of Eq.~(\ref{diag}) thus apparently has to
be a function of all the projection operators \cite{Neumann1929}
\begin{equation}
\label{diag_ens1}
\rho_{\rm{D}}=\sum_{n=1}^D |c(n)|^2 P_n
\end{equation}
which we can also rewrite in exponential form
\begin{equation}
\label{diag_ens}
\rho_{\rm{D}}=\exp(-\sum_{n=1}^D\lambda_nP_n)/Z_{\rm{D}}
\end{equation}
with $Z_{\rm D}=\Tr\exp(-\sum_n\lambda_nP_n)$ and Lagrange multipliers
$\lambda_n$ fulfilling the condition $\langle\Psi_0|P_n|\Psi_0\rangle
=|c(n)|^2=\Tr\{P_n\rho_{\rm D}\}=e^{-\lambda_n}/Z_{\rm D}$.  While it
is easy to check that this ensemble indeed fulfills $\Tr(O\rho_{\rm
  D})= \overline{O}$, see Eq.~(\ref{diag}), by construction and thus
naively proves the ergodic theorem in the quantum case, it depends on
the initial state and fixes the {\it microstate} of the system up to
phase factors which are irrelevant for the long-time dynamics. In
quantum statistical mechanics one is, however, only interested in the
{\it macrostate} consisting of many microstates which cannot be
distinguished by measuring {\it local observables} $O$
\cite{Neumann1929}, see also Fig.~\ref{Schematics}.
While often not sufficiently stressed, this restriction is absolutely
vital. For example, a measurement of any of the non-local projection
operators $P_n$ yields information about the initial state,
$\langle\Psi_0|P_n|\Psi_0\rangle=|c(n)|^2$, and can thus never be
described by a thermal ensemble.

The new important ingredient, which has to be taken into account in
the quantum case, is the distinction between {\it local} and {\it
  non-local} conserved charges. A local operator for a lattice model
is defined as $\mathcal{Q}_n=\sum_{j} q_{j}^n$ where $q_{j}^n$ acts on
lattice sites $j,j+1,\cdots,j+n$ only, with $n$ finite.  In a field
theory this then becomes a fully local operator $\mathcal{Q}_n=\int
\rmd\vec{r}q^n(\vec{r})$.  The number of local conserved charges is
usually finite for a generic quantum system while it increases
linearly---but not exponentially---with system size for an integrable
one-dimensional model \cite{HubbardBook,GrabowskiMathieu}. Using
these definitions, we will in the following consider a system to have
thermalized if and only if, for any {\it local} observable in the TDL,

(i) $\langle O(t\to\infty)\rangle $ becomes time independent and equal to $\overline{O}$, 
and 

(ii)
$\overline{O}=\langle O\rangle_{\rm{th}}=\Tr\{O\rho_{\rm th}\}$ with
$\rho_{\rm th}\!=\exp(-\sum_n \!\beta_n \mathcal{Q}_n)/Z_{\rm th}$
being the appropriate thermal density matrix including all the {\it
  local} conserved charges $\mathcal{Q}_n$ with $Z_{\rm th}$ being the
partition function.  

Here the Lagrange parameters $\{\beta_n\}$ have to be determined by
the set of equations
\begin{equation}
\label{Lagrange_mult}
\langle\Psi_0|\mathcal{Q}_n|\Psi_0\rangle=\Tr\{\mathcal{Q}_n\rho_{\rm
  th}\} 
\end{equation}
because
$\mathcal{Q}_n(t)=\overline{\mathcal{Q}}_n=\langle\Psi_0|\mathcal{Q}_n|\Psi_0\rangle=\mbox{const.}$
\cite{Rigol2007}. Point (ii) is equivalent to the statement that the
reduced density matrix $\rho_{\rm red}$ of the considered subsystem
becomes thermal, $\rho_{\rm red}(t\to\infty)\approx \rho_{\rm th}$
\cite{Goldstein2010,FagottiEssler}. Note that our definition of
thermalization also includes the integrable case where $\rho_{\rm th}$
as defined above is the so-called generalized Gibbs ensemble. This
seems appropriate because the correct statistical ensemble is obtained
the same way in both cases, namely by neglecting the non-local
conserved charges.

In the following we want to show how to get from the trivial
description of the long-time mean of {\it any} operator by the density
matrix $\rho_{\rm D}$ in Eq.~\eqref{diag_ens} to the statistical
description of the long-time mean of {\it local} operators by
$\rho_{\rm th}$ and derive a necessary and sufficient condition under
which such a description is valid. The essential observation is that
we can always replace the set of non-local conserved projection
operators $\{P_1,\cdots, P_D\}$ by
$\{\mathcal{Q}_1,\cdots,\mathcal{Q}_f,P_{f+1},\cdots,P_D\}$, where
$\mathcal{Q}_1,\cdots,\mathcal{Q}_f$ are the $f$-many local conserved
charges of the system, because $[H,\mathcal{Q}_i]=0$ and thus a
representation $\mathcal{Q}_i=\sum_n \mathcal{Q}_i^n P_n$ exists.
Therefore we can write
\begin{equation}
\label{idea}
\rho_{\rm D} \propto
\exp\left(-\sum_{n=1}^f \beta_n\mathcal{Q}_n -\sum_{n=f+1}^D \lambda_n
P_n\right).
\end{equation}
Obviously, $\rho_{\rm D}\to\rho_{\rm th}$ by setting
$\{\lambda_n\}=\{0\}$, i.e., by dropping the non-local conserved
charges from the density matrix.

The rest of our paper is organized as follows: In Sec.~\ref{mazur} we
explain in detail how a rewriting of the observable using the basis
$\{\mathcal{Q}_1,\cdots,\mathcal{Q}_f,P_{f+1},\cdots,P_D\}$ of local
and non-local conserved charges leads to a necessary and sufficient
condition for thermalization. In Sec.~\ref{Sec_finite} we then
investigate this condition numerically using a one-dimensional spin
model as an example. In Sec.~\ref{scaling} we show for the considered
example that the thermalization condition derived here is useful for a
finite-size scaling analysis. In particular, we find that the central
quantity in our thermalization condition shows a clear finite-size
scaling while the fluctuations in $O_{nn}$---contrary to what is
assumed by ETH---seem to stay large. In the final section we conclude.

\section{A Mazur-type equality}\label{mazur}

To obtain a condition for thermalization on the level of matrix
elements of the considered local observable 
we
start from the definition of thermalization, see point (ii) above,
\begin{eqnarray}
\label{condition_simple}
0&=&\lim_{N\to\infty} \left[\,\overline O -\langle O\rangle_{\rm th}\right]\nonumber \\
&=& \lim_{N\to\infty} \sum_{n=1}^{D}  \underbrace{\frac{\langle O  P_n\rangle_{\rm th}}{\langle  P_n^2\rangle_{\rm th}}}_{O_{nn}} \big(\underbrace{\langle \Psi_0| P_n|\Psi_0\rangle}_{|c(n)|^2} -\underbrace{\langle  P_n\rangle_{\rm th}}_{\rho_{\rm th}^{nn}}\big) \nonumber \\
&\stackrel{N\to\infty}{\to}& \int \rmd\varepsilon\; O(\varepsilon)[\underbrace{|c(\varepsilon)|^2\nu(\varepsilon)}_{\Gamma_{\rm ini}(\varepsilon)}-\underbrace{\rho_{\rm th}(\varepsilon)\nu(\varepsilon)}_{\Gamma_{\rm th}(\varepsilon)}]
\end{eqnarray}
where we have used Eq.~\eqref{diag} with the matrix elements $O_{nn}$
rewritten as a thermal expectation value.
In the last line we have, furthermore, introduced a coarse-grained
description in the TDL with $O(\varepsilon)=
\sum_{\varepsilon-d\varepsilon
  <\varepsilon_m<\varepsilon+d\varepsilon} O_{mm}/M_\varepsilon$ where $M_\varepsilon$ is the number of states in the energy interval and, similarly, a
thermal energy distribution $\Gamma_{\rm th}(\varepsilon)$ and
initial energy distribution $\Gamma_{\rm ini}(\varepsilon)$ both
including the coarse-grained density of states $\nu(\varepsilon)$. For
details on the coarse graining see App.~\ref{dist}. The assumption
of the ETH scenario is that $O_{nn}$ becomes a smooth function of the
energy $\varepsilon_n$ in the TDL, i.e.,
$O(\varepsilon=\varepsilon_n)=O_{nn}$ so that a coarse graining is not
required. This poses an unnecessarily restrictive condition. We want
to stress again that it is an essential prerequisite for
thermalization that the initial energy distribution
$\Gamma_{\rm ini}(\varepsilon)$ becomes sharply peaked in the TDL.
Thus most studies concentrate on the question of thermalization after
a quench where this is guaranteed \cite{Rigol2008}.  Since the thermal
distribution $\Gamma_{\rm th}(\varepsilon)$ is also sharply peaked
at the same energy in the TDL by construction, only a small energy
window will contribute to the integral in the last line of
Eq.~(\ref{condition_simple}).

We now derive a thermalization condition by using a Mazur-type
equality to separate $O_{\rm diag}$ into a part proportional to the
local conserved quantities and a part orthogonal to this. A similar
approach has been used previously to understand the role of conserved
charges in quantum transport
\cite{Mazur,Suzuki,ZotosPrelovsek,Prosen,RoschAndrei,SirkerPereira,SirkerPereira2,JungRoschDrude}.
As already briefly explained in the introduction, we create a basis in
operator space made up of two parts instead of using the energy
eigenbasis $P_n=|n\rangle\langle n|$. Firstly, we use the $f$ many
local conserved quantities 
$\widetilde{P}_n\equiv\mathcal{Q}_{n}$
for $n=1,\cdots,f$.
The second part of the basis is composed of $D-f$ many non-local
operators $\widetilde{P}_n$, for $n=f+1,\cdots,D$ such that $\langle
\widetilde P_n\widetilde P_m\rangle_{\rm th} =\langle \widetilde
P_n^2\rangle_{\rm th}\delta_{nm}$ are orthogonal. Here we ask for
orthogonality with respect to the inner product
$\langle\cdots\rangle_{\rm th}=\Tr\{\cdots\rho_{\rm th}\}$ to obtain a
thermalization condition which is trivially fulfilled---as it should
be, see Eq.~(\ref{Lagrange_mult})---if the operator is a linear
combination of the local conserved charges.

To be concrete we consider the case where the system has only a single
relevant local conservation law---the Hamiltonian itself which can be
written as $H=\sum_j \varepsilon_j P_j$ \cite{note4}.
The thermal density
matrix $\rho_{\rm th}=\exp(-H/T)/Z_{\rm th}$ is then just the usual
canonical ensemble expected to describe the system at long times in
the TDL. As the first step we choose the set of normalized operators
$\{H', P'_1, P'_2,\cdots,P'_{D-1}\}$ where $H'=H/\sqrt{\langle
  H^2\rangle_{\rm th}}$, $P'_n=P_n/\sqrt{\langle P_n^2\rangle_{\rm
    th}}$ with $P_n^2=P_n$. The operator
\begin{equation}
\label{Householder1}
U' =(H'-P'_D)/\sqrt{\langle(H'-P'_D)^2\rangle_{\rm th}}
\end{equation}
then defines an orthogonal transformation (a so-called Householder
reflection) for the projection operators
\begin{equation}
\label{Householder3}
\widetilde P_{i+1}=P'_i -2U'\langle U' P'_i\rangle_{\rm th}
\end{equation}
for $i=1,2,\cdots,D-1$ and generates the required orthonormal set
$\{\widetilde P_1\equiv H', \widetilde P_2,\widetilde
P_3,\cdots,\widetilde P_{D}\}$ replacing the projection operators
$\{P_n\}$. With the help of this new basis in operator space we can
split $O_{\rm diag}$ into a `local' and a `non-local' part defined
as follows:
\begin{equation}
\label{Mazur}
O_{\rm diag} \!=\!\sum_n O_{nn}P_n=\underbrace{\frac{\langle O H\rangle_{\rm th}}{\langle H^2\rangle_{\rm th}} H}_{O_{\rm loc}}
+\underbrace{\sum_{n=2}^{D} \frac{\langle O \widetilde P_n\rangle_{\rm th}}{\langle \widetilde P_n^2\rangle_{\rm th}} \widetilde P_n}_{O_{\rm nonloc}}.
\end{equation}
Importantly, energy conservation during time evolution demands
$\overline H=\langle\Psi_0|H|\Psi_0\rangle =\langle H\rangle_{\rm th}$
which fixes the temperature $T$
\cite{Rigol2008,ManmanaWessel,Sedlmayr2013a} and guarantees the
equivalence of the time and canonical ensemble average for the first
term, $O_{\rm loc}$ in Eq.~(\ref{Mazur}), which is proportional to $H$.
With the help of Eqs.~\eqref{diag} and \eqref{Mazur}, the thermalization
condition Eq.~(\ref{condition_simple}) can thus be rewritten as a
condition for the non-local part, $O_{\rm nonloc}$, only.

One is typically interested in expectation values of local observables
$O$, e.g.~correlation functions, which are not affected by an energy
shift $H\to H-E_0$. In this case we can simplify the condition on the
non-local part further by finding an energy shift such that $\langle
O_{\rm nonloc}\rangle_{\rm th}=0$. This is always possible and the
explicit expression for $E_0$ is given in App.~\ref{energyshift_app}.
We stress that this shift is not essential for the following arguments
but rather just simplifies them.
The necessary and sufficient condition for thermalization, which is
one of the main results of this article, then reads
\begin{equation}
\label{condition}
0 =  \lim_{N\to\infty} \overline O_{\rm nonloc} = \lim_{N\to\infty} \sum_{n=2}^{D} 
\underbrace{\frac{\langle O \widetilde P_n\rangle_{\rm th}}{\langle \widetilde P_n^2\rangle_{\rm th}}}_{\widetilde O_{nn}}
\langle \Psi_0|\widetilde P_n|\Psi_0\rangle
\end{equation}
with matrix elements $\widetilde O_{nn}$ defined with respect to the
new operator space basis. Using this basis, the condition for
thermalization now simply states that the time average of the part of
the operator that is a linear combination of the nonlocal conserved
quantities $\widetilde P_2,\cdots,\widetilde P_{D}$ has to vanish, $
\overline O_{\rm nonloc}=0$, while, by construction, $\overline O_{\rm
  loc}\equiv \langle O_{\rm loc}\rangle_{\rm th}$ with $O_{\rm loc}$
being proportional to the local conserved quantity. The thermalization
condition Eq.~(\ref{condition}) allows one to look at fluctuations in matrix
elements of the local observable $O$ in the subspace spanned by the
non-local conserved charges where thermalization takes place instead
of using the energy eigenbasis, on which the ETH is based, which
has no direct relation to the thermalization problem at hand.
Indeed, we show in Sec.~\ref{Sec_finite} that the matrix
elements $\widetilde O_{nn}$ defined in Eq.~(\ref{condition}) show a
completely different finite-size scaling than the matrix elements in
the energy eigenbasis $O_{nn}$. This calculation can be generalized
straightforwardly to the case where many local conservation laws are
present and the condition $ \overline O_{\rm nonloc}=0$ remains
unchanged, see the following subsection.

\subsection{Multiple local conservation laws}

The generalization of \eqref{condition} for a system with $f> 1$ local
conservation laws, $\{\mathcal{Q}_1,\ldots \mathcal{Q}_f\}$, is
straightforward. Note that this includes, in particular, also the case
of integrable lattice models where $f=N$ local charges can be
constructed with $N$ being the number of lattice sites. 

We can still decompose the operator into a local and a non-local part,
\begin{equation}
\label{Mazur2}
O_{\rm diag} =\underbrace{\sum_{n=1}^{f}\frac{\langle O \widetilde P_n\rangle_{\rm th}}{\langle \widetilde
P_n^2\rangle_{\rm th}} \widetilde P_n}_{O_{\rm loc}} +\underbrace{\sum_{n=f+1}^{D} \frac{\langle O \widetilde
P_n\rangle_{\rm th}}{\langle \widetilde P_n^2\rangle_{\rm th}} \widetilde P_n}_{O_{\rm nonloc}}\,,
\end{equation}
with $\{\widetilde P_n\}$ an orthogonal basis set. Firstly for
$n=1,\cdots,f$, $\widetilde{P}_n\equiv\mathcal{Q}_{n}$ are the set of
local conserved quantities.  Secondly we have $\widetilde{P}_n$ with
$n=f+1,\cdots,D$ which are non-local operators defined such that the
set $\{\widetilde P_n\}$ is an orthogonal basis. Such a set can always
be constructed explicitly.

The thermal ensemble average is now given by $\langle
O\rangle_{\rm th}\!=\Tr\{O\rho_{\rm th}\}$ with
$\rho_{\rm th}\!=\exp(-\sum_n \!\beta_n
\mathcal{Q}_n)/Z_{\rm th}$, $Z_{\rm th}=\Tr \exp(-\sum_n
\!\beta_n \mathcal{Q}_n)$.  The Lagrange parameters $\{\beta_n\}$ are
determined by the set of Eqs.~(\ref{Lagrange_mult}),
which in turn ensures that the time and thermal ensemble averages for
$O_{\rm loc}$ are the same by construction, i.e.~$
\overline{O}_{\rm loc}\equiv \langle
O_{\rm loc}\rangle_{\rm th}$ is guaranteed, and the necessary
and sufficient thermalization condition still reads
\begin{equation}
\label{condition2}
\lim_{N\to\infty} \overline O_{\rm nonloc}=0 \, .
\end{equation}
\section{finite-size scaling}
\label{Sec_finite}
Experiments on cold atomic gases as well as most numerical studies of
the thermalization problem are done on finite systems
\cite{Rigol2008,Roux2010,BiroliKollath} where a distinction between
local and non-local conservation laws, strictly speaking, does not
exist. Understanding the scaling with system size of the local and
non-local contributions to the ensemble, Eq.~\eqref{idea}, and the
scaling of the matrix elements $\widetilde O_{nn}$ is therefore a
problem of practical relevance.

So far, our discussion has been general. To test the concepts we have
introduced above we study in the following a specific lattice
model, the one-dimensional anisotropic Heisenberg model,
\begin{eqnarray}
\label{model}
H(\Delta,J_2) &= &J\sum_{j=1}^N h_{j,j+1} +J_2\sum_{j=1}^N h_{j,j+2}\, ,\\\nonumber h_{i,j}& =& \frac{1}{2}\left(S^+_iS^-_{j}+h.c.\right)+\Delta S^z_iS^z_{j}, 
\end{eqnarray}
where $S$ is a spin-$1/2$ operator, $J$ ($J_2$) are the superexchange
couplings for the nearest (next-nearest) neighbors, respectively, and
$\Delta$ parametrizes an exchange anisotropy. In the following we set
$J=1$, use periodic boundary conditions, and study the model by exact
diagonalization
\cite{Konstantinidis} as well as by a light cone renormalization group
(LCRG) algorithm \cite{EnssSirker} for an {\it infinite system}
\cite{Vidal2007}. Here the infinite system size is achieved by
considering transfer matrices in the effective light cone geometry
given by the forward and backward time evolution, see
Ref.~\onlinecite{EnssSirker} and App.~\ref{tdmrg_app} for details. The
model Eq.~(\ref{model}) is integrable for $J_2=0$. In this case the
number of conserved local operators $\mathcal{Q}_n$ increases linearly
with system size $N$. In the non-integrable case $J_2\neq 0$, $H$
itself and $S^z_{\rm{tot}}=\sum_j S^z_j$ are the only conserved local
operators. In this paper we will only discuss the generic,
non-integrable case $J_2\neq 0$. To guarantee that the initial energy
distribution becomes sharply peaked in the TDL, we consider a quench
scenario \cite{BarmettlerPunk,BarmettlerPunk2}. As the initial state we
choose the ground state $|\Psi_0(\Delta,J_2)\rangle$ of the
Hamiltonian Eq.~(\ref{model}) with parameters $\Delta$ and $J_2$.  We
then time evolve with $H(\Delta',J'_2)$ where $(\Delta',J'_2)\neq
(\Delta,J_2)$.  Because $\langle\Psi_0|S^z_{\rm
  tot}|\Psi_0\rangle\equiv 0$ for zero magnetic field the associated
Lagrange multiplier is zero as well and $\rho_{\rm th}=\exp(-H/T)/Z$.

\subsection{Locality and statistical ensembles}

\begin{figure}
\begin{center}
\includegraphics*[width=0.95\columnwidth]{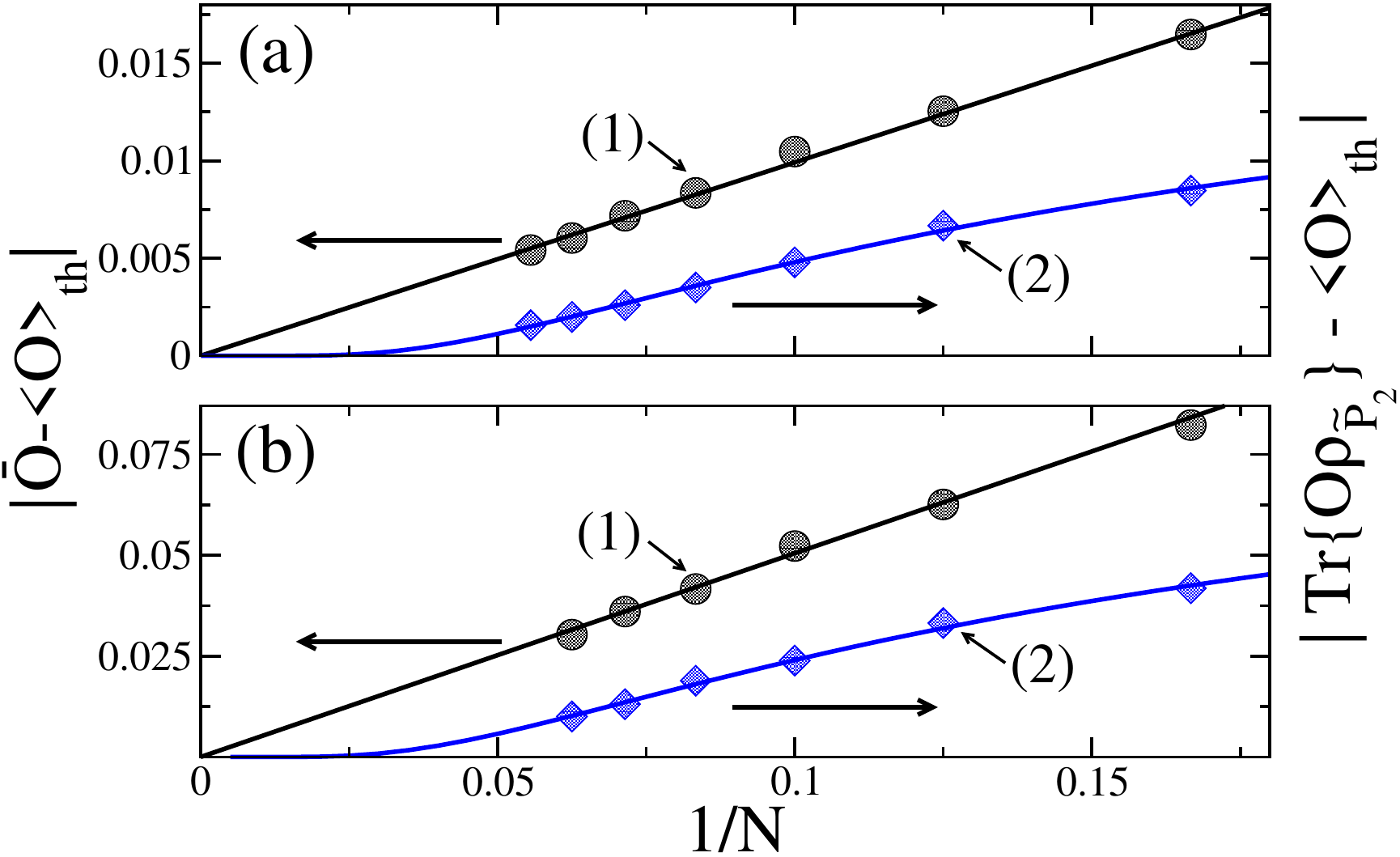}
\end{center}
\caption{Difference of time and ensemble averages, following a quench
  in an $N$-site system.  (a) Quench with $|\Psi_0(5,0.2)\rangle$,
  $H(1,0.2)$ and $O=\vec{S}_i\cdot\vec{S}_{i+1}$ with (1) $|\overline O -
  \langle O\rangle_{\rm th}|$ and a fit $\sim 0.1/N$, and (2)
$|\Tr\{O\rho_{\widetilde P_2}\}-\langle O\rangle_{\rm th}|$, see Eq.~\eqref{extended}, and a fit $\sim 0.02\exp(-0.14 N)$. Symbols denote
  the exact diagonalization data; lines are the fits. (b) As for
  (a) but with $O=\vec{S}_i\cdot\vec{S}_{i+2}$ and fits (1) $\sim 0.5/N$,
  and (2) $\sim 0.1\exp(-0.14 N)$. }
\label{Fig1}
\end{figure}

We start by investigating the step from $\rho_D$ to $\rho_{\rm th}$,
Eq.~\eqref{idea}, i.e., ignoring the contributions of the non-local
conserved charges to the statistical ensemble. More specifically, we
want to study how much keeping one of the non-local charges in the
density matrix affects the expectation values of local operators as a
function of the system size. To do so we define an extended canonical
ensemble,
\begin{equation}
\label{extended}
\rho_{\widetilde P_j} =\exp(-\beta H + \lambda_j \widetilde P_j)/Z_{\widetilde P_j},
\end{equation}
with $\widetilde P_j$ being a non-local conserved charge as defined in
Eq.~(\ref{Householder3}), $Z_{\widetilde P_j}=\Tr\exp(-\beta H +
\lambda_j \widetilde P_j)$, and the two parameters $\beta$ and $\lambda_j$
being determined by the conditions $\overline
H=\langle\Psi_0|H|\Psi_0\rangle =\Tr\{H\rho_{\widetilde P_j}\}$ and
$\overline{\widetilde P}_j=\langle\Psi_0|\widetilde P_j|\Psi_0\rangle
=\Tr\{\widetilde P_j\rho_{\widetilde P_j}\}$. This extended ensemble
is compared with
the canonical ensemble in Fig.~\ref{Fig1}(a,b) (data denoted by
diamonds) using $\widetilde P_2$ as a generic example.
Indeed, we find that the qualitative results for the finite-size
scaling are independent of which of the non-local $\widetilde P_j$,
$j=2,\cdots,D$ we add in the extended ensemble Eq.~(\ref{extended})
and consistent with the following: (I) The change in the average of a
{\it local} observable caused by including an additional non-local
conservation law vanishes exponentially with system size,
i.e.~$|\Tr\{O\rho_{\widetilde P_j}\}-\langle O\rangle_{\rm th}|\sim
\langle O\rangle_{\rm th}\e^{-N}$. This is corroborated by the
excellent agreement between the exponential fits and the
diagonalization data in Fig.~\ref{Fig1}. (II) The contribution of all
non-local conserved quantities to the ensemble average of a local
observable vanishes linearly in $1/N$, i.e.~$|\overline O - \langle
O\rangle_{\rm th}|\sim \langle O\rangle_{\rm th}/N$ (see data denoted
by circles in Fig.~\ref{Fig1}(a,b)).  This follows from the fact that
$\overline O\equiv\Tr\{ O\rho_{\rm D}\}$ and the canonical ensemble
$\rho_{\rm th}$ is obtained from $\rho_{\rm D}$, see Eq.~(\ref{idea}),
by neglecting all non-local conserved charges.  Again, the linear fits
in Fig.~\ref{Fig1} clearly support this statement in the considered
example.  From the data presented in Fig.~\ref{Fig1} we see, however,
that for a finite system the non-local conserved quantities do
contribute, showing that the TDL is essential for a full
thermalization.

\subsection{Locality and observables}

\begin{figure}
\begin{center}
\includegraphics*[width=0.95\columnwidth]{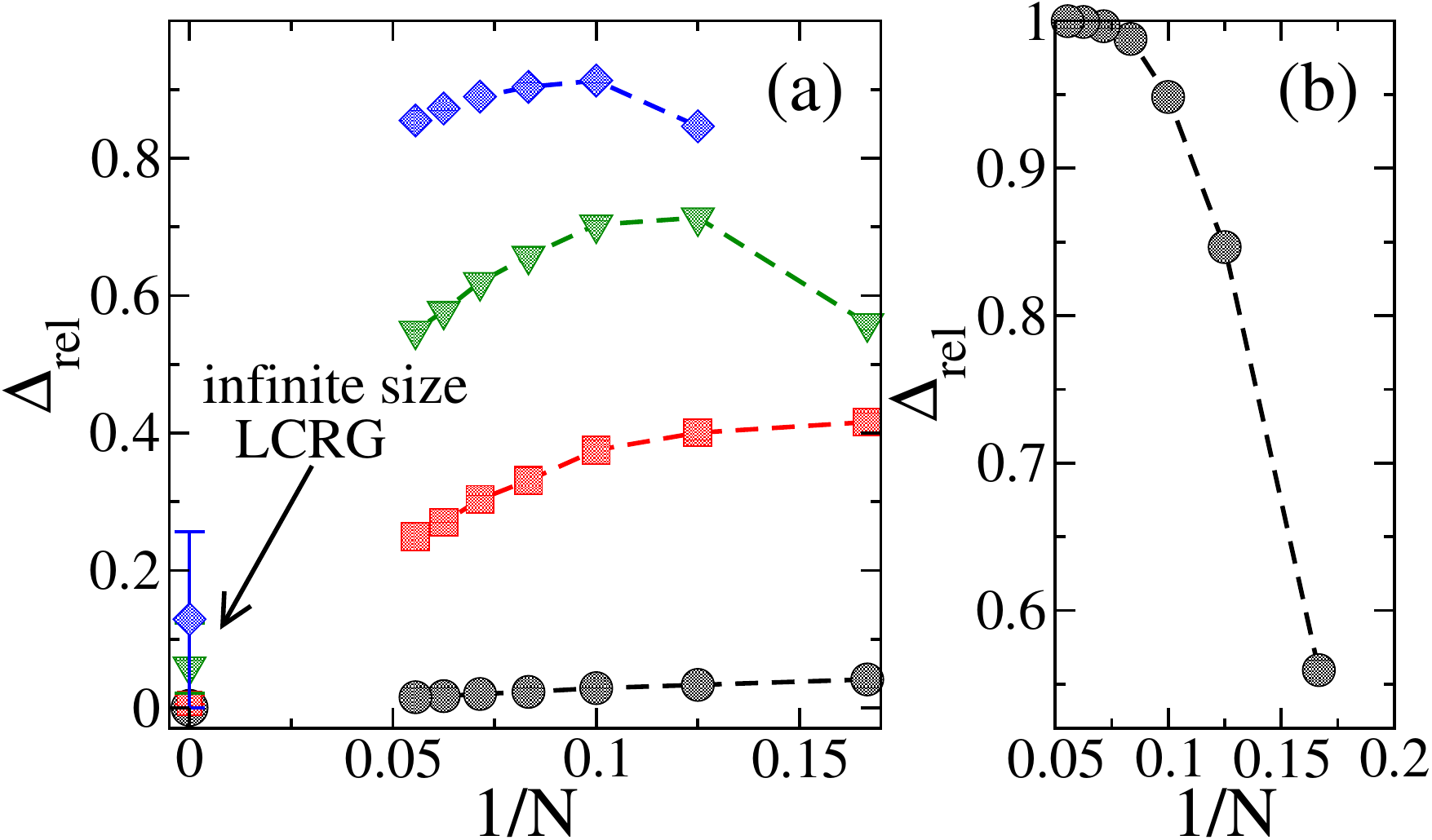}
\end{center}
\caption{
  Scaling of the relative
  deviation
  $\Delta_{\rm rel}$, Eq.~(\ref{rel_error}), for
  $O=\vec{S}_i\cdot\vec{S}_j$ with (a) $|i-j|=1,2,3,4$ (from bottom to
  top) and (b) $|i-j|=N/2$. The TDL data in (a) from LCRG, see
  App.~\ref{tdmrg_app} for details, are consistent with thermalization
  within error bars.}
\label{Fig3}
\end{figure}

Next, we want to study how the amount of locality of the operator
itself affects its thermalization. As an example we consider again the
quench with $|\Psi_0(5,0.2)\rangle$ and $H(1,0.2)$ as in
Fig.~\ref{Fig1}. Numerical data for $O=\vec{S}_i\cdot\vec{S}_j$ are shown
in Fig.~\ref{Fig3}.
The relative deviation
\begin{equation}
\label{rel_error}
\Delta_{\rm rel}=\left|\frac{\overline{\vec{S}_i\cdot\vec{S}_j}-\langle\vec{S}_i\cdot\vec{S}_j\rangle_{\rm{th}}}{\overline{\vec{S}_i\cdot\vec{S}_j}}\right|
\end{equation}
between the time and the canonical ensemble average for finite systems
becomes larger the larger the distance is and thus the less local $O$
is, see Fig.~\ref{Fig3}(a). If we fix the distance between the spin
operators to $N/2$ as in Fig.~\ref{Fig3}(b) then the canonical
ensemble average approaches zero with increasing $N$ much faster than
the time average so that $\Delta_{\rm rel}\to 1$. The numerical
results thus support a picture of a finite subsystem which
thermalizes with the rest of the quantum system acting as an effective bath in the TDL, as shown schematically in
Fig.~\ref{Schematics}.
Note that the data in Fig.~\ref{Fig3}(a) obtained by LCRG clearly
support thermalization in the TDL within the error bars, which stem from
approximating the initial state, a discrete time evolution, and the
finite simulation time, see App.~\ref{tdmrg_app}.

\subsection{Projection onto locally conserved charges}

\begin{figure}
\begin{center}
\includegraphics*[width=0.95\columnwidth]{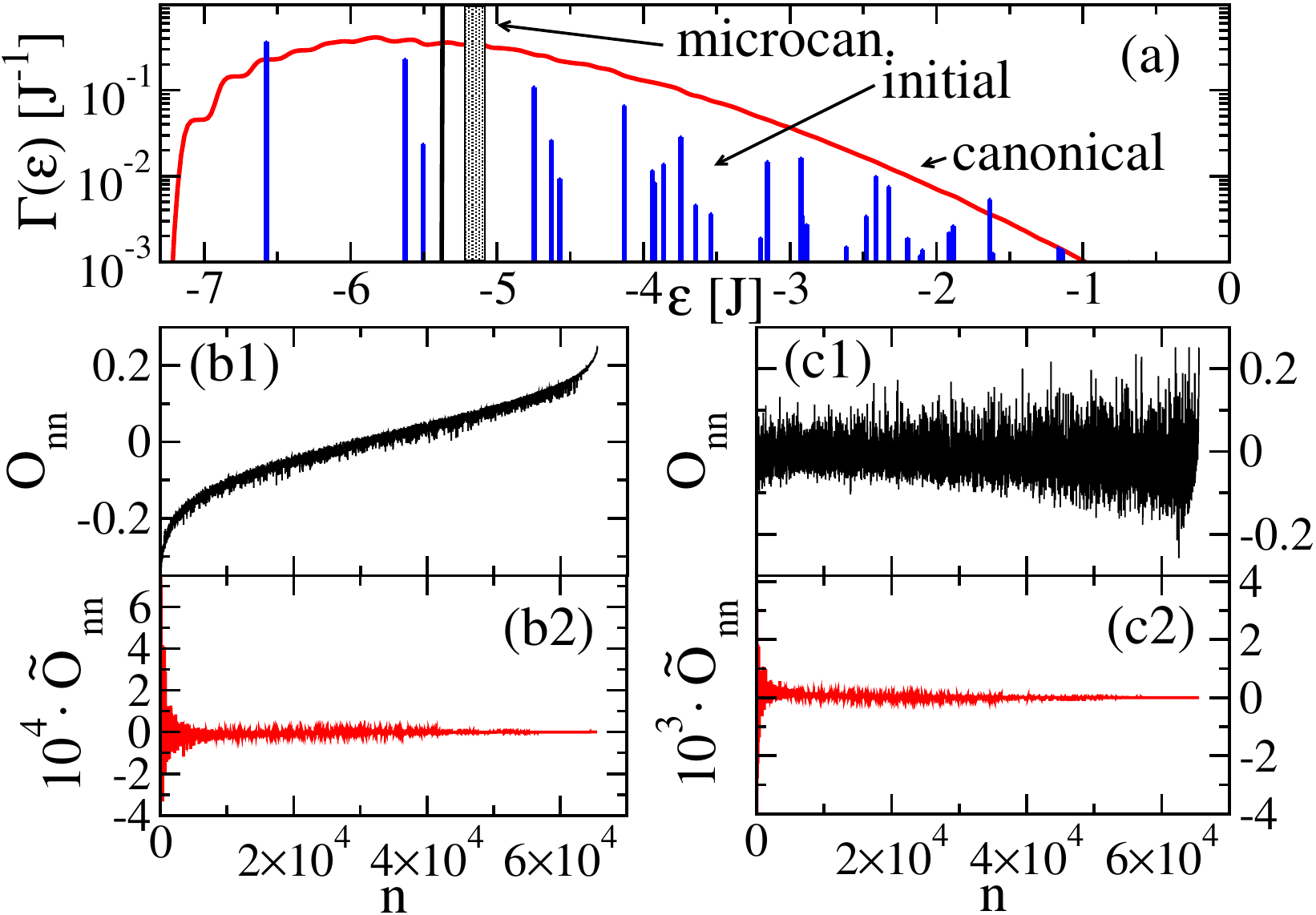}
\end{center}
\caption{Same quench as in Fig.~\ref{Fig1}, for $N=16$.  
  (a) Initial, microcanonical and canonical energy distribution
  functions. Results for (b) $O=\vec{S}_i\cdot\vec{S}_{i+1}$ and (c)
  $O=\vec{S}_i\cdot\vec{S}_{i+4}$. (b1) and (c1) show $O_{nn}$ while
  (b2) and (c2) show $\widetilde O_{nn}$. Note the different $y$-axis scale in (b2) and (c2).}
\label{Fig4}
\end{figure}
The initial distribution, $|c(n)|^2$, and the microcanonical and
canonical ones, $\Gamma_{\rm{mic/can}}(\varepsilon)$, are shown in
Fig.~\ref{Fig4}(a), for the same quench considered in
Figs.~\ref{Fig1} and \ref{Fig3}. While the initial distribution will
become singly peaked in the TDL \cite{Rigol2008}, this is clearly not
the case for the system sizes within reach of exact diagonalization.
The initial state distribution thus cannot be simply replaced by the
microcanonical ensemble.  Furthermore, the matrix elements of the
local operator in the energy eigenbasis of the time evolving
Hamiltonian, $O_{nn}$, shown in Figs.~\ref{Fig4}(b1) and \ref{Fig4}(c1) for two
different correlation functions, show large fluctuations and, as shown
in Sec.~\ref{scaling}, no clear finite-size scaling. Thus our data do
not support the ETH scenario---at least not for the considered system
sizes---yet we already see clear indications that subsystems will
thermalize in the TDL as shown in Figs.~\ref{Fig1} and \ref{Fig3}.

To understand these findings we
return to the
necessary and sufficient condition for thermalization,
Eq.~(\ref{condition}), which requires that the time average of the
non-local part vanishes.  Indeed, the matrix elements
$\widetilde{O}_{nn}$ in the subspace spanned by the non-local
conserved charges, shown in Figs.~\ref{Fig4}(b2) and (c2), have
fluctuations centered around zero which are orders of magnitude
smaller than those in $O_{nn}$.
Furthermore, the fluctuations in $\widetilde O_{nn}$ show a clear
finite-size scaling with $\widetilde O_{nn}\to 0$ apparently
exponentially for $N\to\infty$, see Fig.~\ref{Fig5}(a) and the
discussion in Sec.~\ref{scaling}, while no such scaling is seen for
the fluctuations in the matrix elements $O_{nn}$ in the energy
eigenbasis.

\section{Fluctuations in $O_{nn}$ and $\widetilde O_{nn}$ and the
  eigenstate thermalization hypothesis}\label{scaling}

According to the ETH, $O_{nn}$ should
become a smooth function of the eigenenergies $\varepsilon_n$ in the
TDL. For the system sizes we are able to exactly
diagonalize we are clearly far from that limit and fluctuations in
$O_{nn}$ are large, see Fig.~\ref{Fig4}(c1). Nevertheless, we can
still check how these fluctuations scale with system size.

In order to investigate the scaling with system size we define the
average size of the fluctuations in an energy interval:
\begin{equation}\label{Onn_scaling}
\Delta_O=\sum_{n=1}^D\Gamma(\varepsilon_n-E)\left|O_{nn}-\left<O_{nn}\right>_{mc}\right|\,.
\end{equation}
$\Gamma(\varepsilon_n-E)$ restricts the sum to to an energy interval of $0.05$ times the bandwidth $W_\Delta=\varepsilon_D-\varepsilon_1$, and
centered on the middle of the spectrum, $E=W_\Delta/2+\varepsilon_1$, with $\varepsilon_1$ the ground state energy. I.e.
\begin{equation}
\Gamma(\varepsilon)=\frac{1}{M_\varepsilon}\left(\Theta\left[\varepsilon+0.05W_\Delta\right]-\Theta\left[\varepsilon-0.05W_\Delta\right]\right)
\end{equation}
with $\Theta(\varepsilon)$ the Heaviside function and $M_\varepsilon$ the number of states in the energy interval.
 $\left<O_{nn}\right>_{mc}$ is the locally defined average, in other words the microcanonical ensemble, calculated here with the same energy window
\begin{equation}
\left<O_{nn}\right>_{mc}=\sum_{m=1}^DO_{mm}\Gamma(\varepsilon_m-\varepsilon_n)\,.
\end{equation}
To compare the size of the fluctuations with the magnitude of the
operator we define $\left<O\right>_{E}=\sum_n\Gamma(\varepsilon_n-E)O_{nn}$.  We can use the
same definition to study fluctuations in $\widetilde O_{nn}$, which
are the matrix elements appearing in the thermalization condition,
Eq.~\eqref{condition}, after the basis rotation, by writing
\begin{equation}
\Delta_{\widetilde{O}}=\sum_{n=1}^D\Gamma(\varepsilon_n-E)\left|\widetilde{O}_{nn}-\left<\widetilde{O}_{nn}\right>_{mc}\right|
\end{equation}
with the interval for the sum defined as for Eq.~\eqref{Onn_scaling}. Strictly speaking $\left<\widetilde{O}_{nn}\right>_{mc}$ is
no longer the microcanonical ensemble average as $n$ no longer labels the eigenenergies.  Nonetheless one can define an analog and we retain
the same notation for ease of presentation.

\begin{figure}
\begin{center}
\includegraphics*[width=0.95\columnwidth]{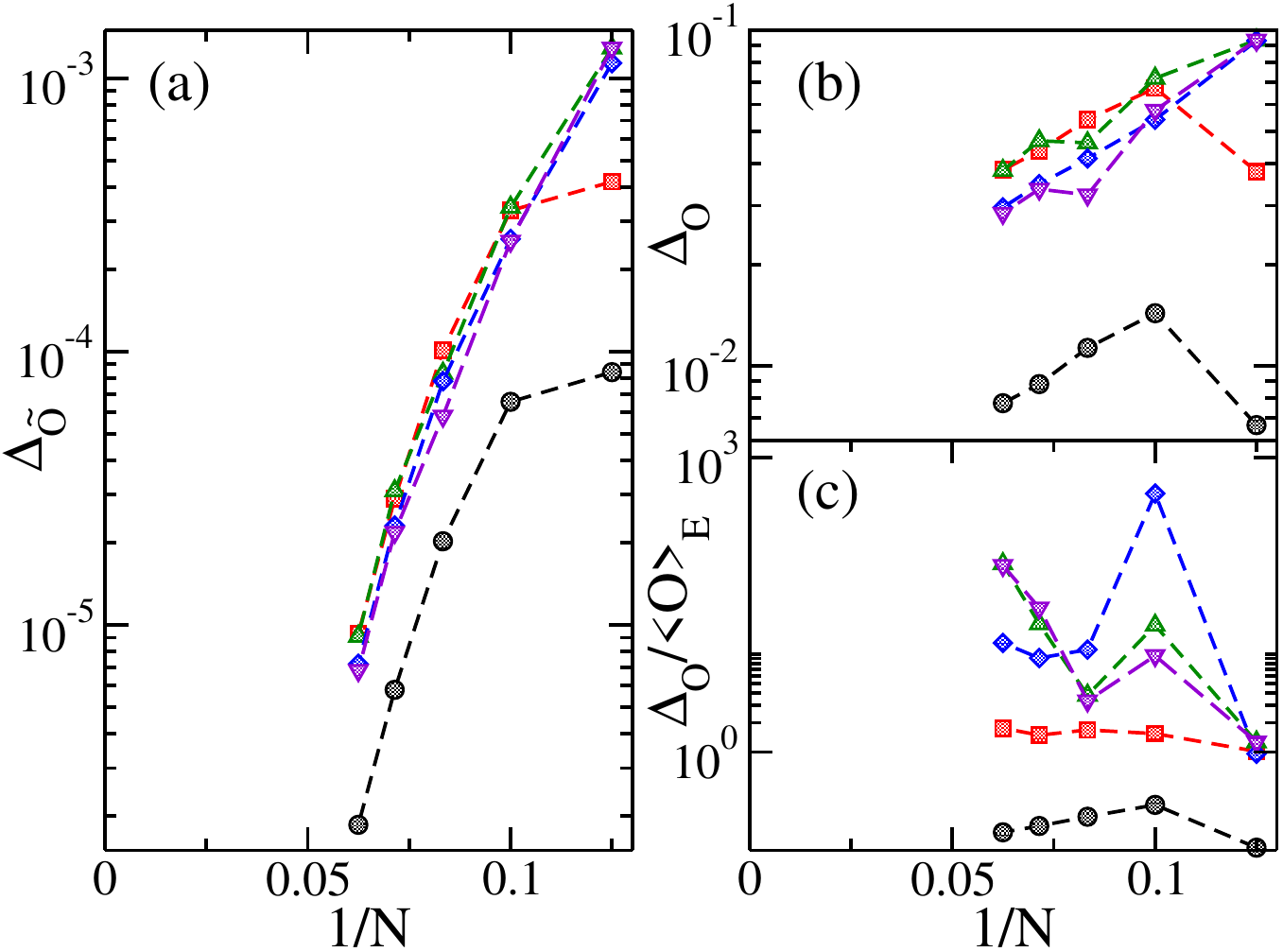}
\end{center}
\caption{Quench with $|\Psi_0(5,0.2)\rangle$ and $H(1,0.2)$. Comparison of (a) $\Delta_{\widetilde{O}}$, (b) $\Delta_O$, and
  (c) $\Delta_O/\langle O\rangle_E$ for different system sizes from
  $N=8$ to $16$. Plotted are the observables $O=\vec{S}_i\cdot\vec{S}_{j}$
  with $|i-j|=1$ (black circles), $|i-j|=2$ (red squares), $|i-j|=3$
  (upward green triangles), $|i-j|=4$ (blue diamonds), and $|i-j|=5$ (downward purple
  triangles). $\Delta_{\widetilde{O}}$ shows clear exponential scaling
  to zero (note the logarithmic scales).}
\label{Fig5}
\end{figure}

We consider again the same quench as in Fig.~\ref{Fig1} with
$|\Psi_0(5,0.2)\rangle$ and $H(1,0.2)$, and look at observables
$O=\vec{S}_i\cdot\vec{S}_{j}$ for different $|i-j|$. In Fig.~\ref{Fig5}
we plot $\Delta_O$ and $\Delta_{\widetilde{O}}$ for system sizes $N=8$
to $16$. The absolute magnitude of the fluctuations of $\Delta_O$,
Fig.\ref{Fig5}(b), can be several orders of magnitude larger than
their average value even for $N=16$, see Fig.~\ref{Fig5}(c). In
particular, no clear-cut scaling with system size can be seen. The
relative size of the fluctuations, defined as $\Delta_O/\langle
O\rangle_{E}$ as in Ref.~\onlinecite{Rigol2009}, even seems to increase with the
system size for distances $|i-j|\geq 2$, see Fig.~\ref{Fig5}(c).  We
thus have to conclude that, at least for the achievable system sizes, we
do not see any indications for $O_{nn}$ to become a smooth function as
assumed by the ETH, while we do
already see clear indications that the system will eventually
thermalize, see Figs.~\ref{Fig1} and \ref{Fig3}(a).

If we consider, on the other hand, the matrix elements $\widetilde
O_{nn}$ in the thermalization condition, Eq.~(\ref{condition}), we do
see a clear scaling to zero, which seems to depend exponentially on
the system size, see Fig.~\ref{Fig5}(a). So while the fluctuations in
$O_{nn}$ are large, the matrix elements in the appropriate basis,
$\widetilde{O}_{nn}$, are seen to rapidly decrease already for the
small system sizes considered here underlining the usefulness of the
necessary and sufficient condition (\ref{condition}) to investigate
thermalization by a finite-size scaling analysis of numerical data.
This condition also provides an alternative explanation compared to
the ETH of why thermalization is independent from the initial state as
long as the initial distribution function $|c(n)|^2$ becomes sharply
peaked in the TDL. As already noted, condition
(\ref{condition}) seems to depend explicitly on the initial state.
However, if $\widetilde O_{nn}\sim \e^{-N}$ and
$\langle\Psi_0|\widetilde P_n|\Psi_0\rangle \approx
\langle\Psi_0|P_n|\Psi_0\rangle$, see App.~\ref{rotation_app}, is
sharply peaked in the TDL then only a limited number of terms
contribute and the thermalization condition is fulfilled independent
of the exact form of the initial distribution.

\section{Conclusions}

To summarize, we have looked at the question of thermalization in
closed quantum systems from the perspective of conservation laws.
Contrary to a classical system two distinct kinds are always present:
local and non-local ones. While the diagonal ensemble $\rho_D$,
Eq.~(\ref{diag_ens1}), is a function of all local and non-local
conservation laws, depends explicitly on the initial state, and
describes the time average of {\it any} observable in a finite system
by construction, the thermal ensemble $\rho_{\rm th}$ is a function of
the local conserved charges only. This distinction between local and
non-local charges also makes it clear why $\rho_{\rm th}$ is the
microcanonical or canonical ensemble for generic quantum systems while
it becomes a generalized Gibbs ensemble for integrable quantum systems
which have an extensive number of local conservation laws. Contrary to
the diagonal ensemble, only the time averages of {\it local}
observables in the TDL can be replaced by a
statistical average using $\rho_{\rm th}$, i.e., only a reduced
density matrix $\rho_{\rm red}$ for a subsystem can fulfill
$\lim_{t\to\infty}\lim_{N\to\infty} \rho_{\rm red}(t)=\rho_{\rm th}$
\cite{note5} while the density matrix for the whole system describes a
pure state at all times.
Furthermore, local observables will thermalize
differently depending on how large their overlap with the local
conserved charges is. To make this explicit, we have derived an
equivalent thermalization condition separating the observable into a
projection onto the local conserved charges, for which the time and
statistical average agree by construction, and a projection onto the
non-local ones where thermalization takes place. This thermalization
condition will, in particular, be trivially fulfilled if the
observable is a linear combination of the local conserved charges.

Importantly, the condition is strict and not based on the ETH. In
particular, it explicitly shows that systems can exist which do
thermalize although the ETH does not apply. One possible example of
such a system is the spin-$1/2$ system which we numerically
investigated in the second part of our paper. We find strong numerical
indications that the system will thermalize in the TDL while the
fluctuations in the matrix elements $O_{nn}$ do not seem to show a
clear finite-size scaling as would be expected by ETH. The
fluctuations in the modified matrix elements $\widetilde O_{nn}$,
obtained by using instead an operator space basis consisting of the
local and non-local conserved charges, do show, on the other hand,
exponential scaling to zero, strongly supporting the usefulness of the
thermalization condition derived in this work. Furthermore, the
exponential scaling implies that the thermalization condition,
Eq.~(\ref{condition}), is independent of the precise initial distribution
as long as this distribution becomes sharply peaked in the TDL.
Equation (\ref{condition}) therefore provides a general alternative
thermalization scenario.

\acknowledgments The authors thank F.H.L.~Essler, B.~Fine,
A.~Polkovnikov, and M.~Rigol for discussions and acknowledge support
from the Collaborative Research Centre SFB/TR49 and the Graduate School
of Excellence MAINZ. We are grateful to the Regional Computing Center
at the University of Kaiserslautern and the AHRP for providing
computational resources and support.

\appendix

\section{The energy shift}\label{energyshift_app}

We are interested in expectation values of observables which are
independent of a shift in energy.  By shifting the energy, $H\to
H-E_0$, the projection operators $\widetilde P_n$ are modified
because of the orthogonality condition $\langle H\widetilde
P_n\rangle=0$. The qualitative behavior of $\widetilde{O}_{nn}$ is,
however, not affected. A convenient, unique gauge is obtained by
demanding that $\langle O_{\rm nonloc}\rangle_{\rm th}=0$. Focusing
once again on a system with $H$ being the only local conserved quantity,
this is achieved by choosing
\begin{equation}
E_0=\frac{\langle OH\rangle_{\rm th}\langle H\rangle_{\rm th}-\langle O\rangle_{\rm th}\langle H^2\rangle_{\rm th}}{\langle OH\rangle_{\rm th}-\langle O\rangle_{\rm th}\langle H\rangle_{th}}\,,
\end{equation}
which is the shift we have used in the text.
For a system with $f> 1$ local conserved quantities a similar condition can be found.

\section{Relation between the original and the rotated basis}\label{rotation_app}

The relation between the old and the new operator basis can
be expressed as
\begin{eqnarray}
\widetilde P_{m+1}&=&\sum_{n=1}^D a_n^{m+1} P_n\\\nonumber&=&\sum_{n=1}^D \underbrace{a_n^{m+1} \sqrt{\langle P_n\rangle_{\rm th}}}_{\langle\widetilde P_{m+1} P'_n\rangle_{\rm th}} P'_n \; (m=1,\cdots,D-1)
\end{eqnarray}
where $P'_n =P_n/\sqrt{\langle P_n^2\rangle_{\rm th}}$ are the normalized projection operators, $\langle P'_i P'_j\rangle_{\rm th}=\delta_{ij}$ and
$P_n^2=P_n$. Using the definition, Eq.~\eqref{Householder3}, of the Householder reflection and some simple algebra we find
\begin{equation}
\label{coeffs}
\langle\widetilde P_{m+1}P'_n\rangle_{\rm th}=\delta_{nm}-\frac{2\varepsilon_n\varepsilon_m}{\langle(H'-P'_D)^2\rangle_{\rm th}}
\frac{\sqrt{\langle P_n\rangle_{\rm th}}\sqrt{\langle P_m\rangle_{\rm th}}}{\langle H^2\rangle_{\rm th}}.
\end{equation}
While $\langle(\ H'-P'_D)^2\rangle_{\rm th}\propto\mathcal{O}(1)$
and $\langle H^2\rangle\propto \mathcal{O}(N^2)$, we have $\langle
P_n\rangle_{\rm th}\propto \e^{-N}$ so that the expansion
coefficients $a_n^{m+1}$ are sharply peaked at $n=m$. As a consequence,
the initial distribution is not affected by the rotation in the
TDL, i.e.,
\begin{equation}
\langle\Psi_0|\widetilde P_{n+1}|\Psi_0\rangle \stackrel{N\to\infty}{\to} \langle\Psi_0|P'_n|\Psi_0\rangle \quad (n=1,\cdots, D-1)
\end{equation}
and becomes sharply peaked in the TDL. The matrix
elements of the observable
\begin{eqnarray}
\label{Bar_O}
\widetilde O_{mm}&=&\frac{\langle O\widetilde P_m\rangle_{\rm th}}{\langle \widetilde P_m^2\rangle_{\rm th}}=\sum_n O_{nn} \langle P_n\widetilde P_m\rangle_{\rm th} \nonumber \\
&=&\sum_n a_n^m\langle P_n\rangle_{\rm th} O_{nn}
\end{eqnarray}
are, however, changed because they are given by summing over the
exponentially many matrix elements $O_{nn}$ so that the exponentially
small corrections in $a_n^m$, see Eq.~(\ref{coeffs}), still matter.

\section{Energy distributions and coarse graining}\label{dist}

\begin{figure}
\begin{center}
\includegraphics*[width=0.95\columnwidth]{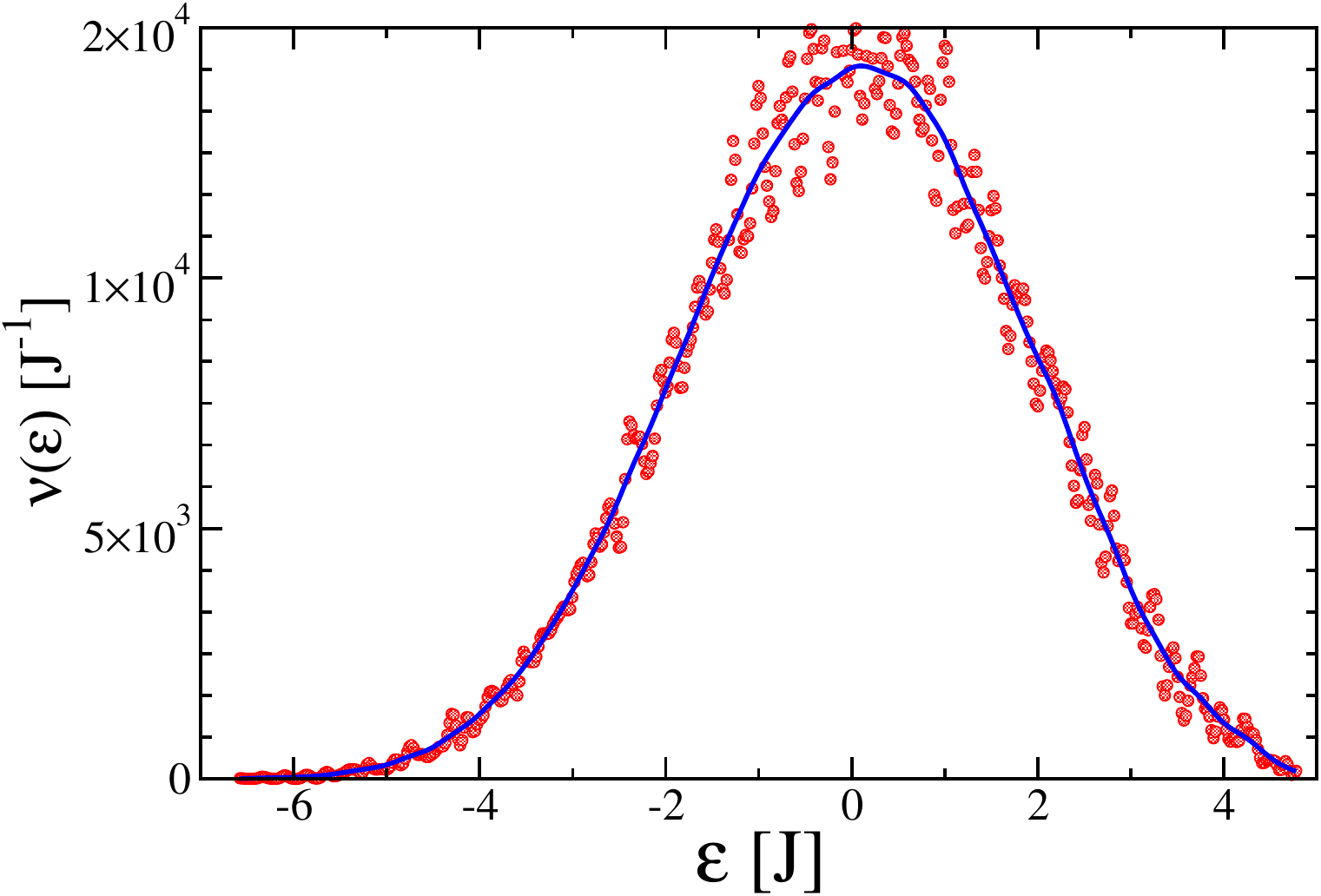}
\end{center}
\caption{Coarse grained density of states, $\nu(\varepsilon)$, for the
  Hamiltonian $H(1,0.2)$ with $N=16$. The coarse graining width is
  $W=10\delta$, where $\delta$ is the mean level spacing. Shown are the
  result after coarse graining (red circles), and the result after an
  additional running average (blue line).}
\label{Fig1_App}
\end{figure}

In order to plot the continuum energy distributions a coarse graining is necessary. The density of states is first made continuous by approximating
\begin{equation}
\nu(\varepsilon)\equiv\sum_n\delta(\varepsilon-\varepsilon_n)\approx\sum_n\chi_{W}(\varepsilon-\varepsilon_n)\,,
\end{equation}
with an envelope function:
\begin{equation}
\chi_{W}(\varepsilon)=\frac{e^{-\varepsilon^2/(2W^2)}}{\sqrt{2 \pi W^2}}\,.
\end{equation}
In this paper we have used $W=10\delta$ for $N=16$, where $\delta$ is
the mean level spacing with an additional running average. The
results of these procedures for the density of states are shown in Fig.~\ref{Fig1_App}. The same procedure is
performed for the canonical ensemble. As a check that this is working
correctly one must compare operator averages found with these coarse
grained distributions and with the exact ones. Note that whilst a
coarse graining over a wider energy range (\emph{e.g.}~$W=50\delta$)
will give the same result for the density of states as in
Fig.~\ref{Fig1_App}, it does not give accurate results for the
canonical ensemble.
For the microcanonical ensemble one simply broadens the delta-function
around the initial energy $E=\langle\Psi_0|H|\Psi_0\rangle$,
\begin{equation}
\Gamma_{\rm mic}(\varepsilon)\approx\frac{1}{M_\varepsilon}\left(\Theta(\varepsilon-E+W/2)-\Theta(\varepsilon-E-W/2)\right)\,,
\end{equation}
where, again, $\Theta(\varepsilon)$ is the Heaviside function and $M_\varepsilon$ is the number of states in the energy interval.

In principle one could also attempt this procedure on the initial
distribution to calculate the time average. However, for the system
sizes we are able to consider we find that it is not possible to
smoothen the initial distribution and, at the same time, retain
accurate averages for physical quantities.

\section{Light cone renormalization group}\label{tdmrg_app}

\begin{figure}
\begin{center}
\includegraphics*[width=0.95\columnwidth]{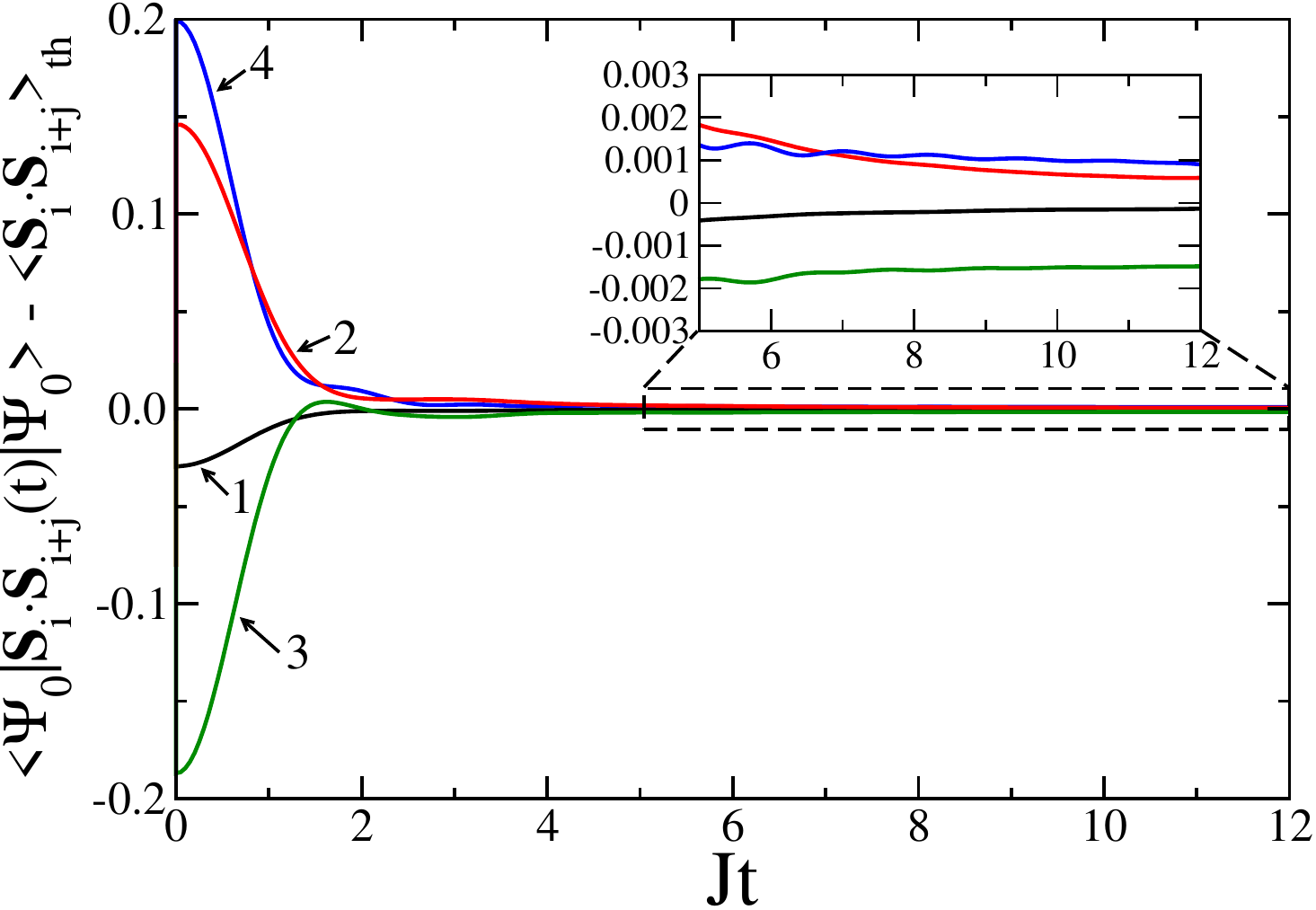}
\end{center}
\caption{Difference between the time dependent expectation value after
  the quench considered in the main text and the thermal expectation
  value,
  $\langle\Psi_0|\vec{S}_i\cdot\vec{S}_{i+j}(t)|\Psi_0\rangle-\langle\vec{S}_i\cdot\vec{S}_{i+j}\rangle_{\rm{th}}$,
  for distances $j=1,2,3,4$ as indicated. Inset: Behavior at the longest simulation times in more detail.}
\label{Fig2_App}
\end{figure}

In order to show that the quench considered in Fig.~\ref{Fig1} and
Fig.~\ref{Fig3} does indeed lead to a thermalization of the local
correlation functions at long times in the TDL we have
performed time-dependent DMRG calculations for infinite system size.
The time evolution is performed by using a third order Trotter-Suzuki
decomposition with a time step $J\delta t=0.05$.  In order to obtain
results in the TDL we have simulated the dynamics on a
light cone which grows with an effective velocity, set by the Trotter
time step, which is much larger than the Lieb-Robinson velocity.
Further details of the algorithm are given in Ref.~\onlinecite{EnssSirker}. The
initial state and the thermalized state were calculated using
an imaginary time evolution.

In Fig.~\ref{Fig2_App} we show results for
$\langle\Psi_0|\vec{S}_i\cdot\vec{S}_{i+j}(t)|\Psi_0\rangle-{\langle\vec{S}_i\cdot\vec{S}_{i+j}\rangle_{\rm{th}}}$
for the quench with $|\Psi_0(5,0.2)\rangle$ and $H(1,0.2)$ considered
in the main paper.
At the longest times we can simulate, this difference is of order
$10^{-3}$. Due to the Trotter decomposition we expect an error of
order $(\delta t)^2\sim 10^{-3}$ so that the system has already
thermalized within error bars. While we could, in principle, reduce
the Trotter step $\delta t$ we also see that a full equilibration has
not taken place yet so that a tighter bound on thermalization would in
addition require substantially longer simulation times which are not
feasible using present day computers and algorithms. Note that the
relative deviation $\Delta_{\rm rel}$ shown in Fig.~\ref{Fig3}(a) is extremely sensitive to small errors because the
difference plotted in Fig.~\ref{Fig2_App} is divided by the time
average of the correlator. For the longer-range correlation functions
this value becomes very small---we obtain, for example,
$\overline{\vec{S}_i\cdot\vec{S}_{i+4}}\approx 0.007$---thus magnifying the
numerical error in the time-dependent correlation function.


\begin{thebibliography}{48}
\bibitem[{\citenamefont{Footnote}(2014)}]{note1}\bibinfo{note}{One still has to assume that the initial state overlaps with a macroscopic number of energy eigenstates of the time evolving Hamiltonian in the TDL.}
\bibitem[{\citenamefont{Footnote}(2014)}]{note2}\bibinfo{note}{Position space, however, is by no means special: For a Hamiltonian with e.g.~local interactions in momentum space the same approach can be used and the results directly carry over.}
\bibitem[{\citenamefont{Footnote}(2014)}]{note3}\bibinfo{note}{In the case of degeneracies, Eq.~\eqref{diag} remains valid if we choose the eigenbasis in which $O$ is diagonal in the degenerate subspaces.}
\bibitem[{\citenamefont{Footnote}(2014)}]{note4}\bibinfo{note}{Local quantities like the momentum, angular momentum, or particle number are also often conserved. If their expectation value in the initial state is zero, however, then the corresponding Lagrange multiplier is zero as well, see Eq.~\eqref{Lagrange_mult}, so that these local charges do not show up explicitly in the density matrix.}
\bibitem[{\citenamefont{Footnote}(2014)}]{note5}\bibinfo{note}{We want to remind the reader here that we have assumed that $\protect \qopname  \relax m{lim}_{t\to \infty }\protect \qopname  \relax m{lim}_{N\to \infty } \rho _{\protect \rm  red}(t)=\protect \qopname  \relax m{lim}_{N\to \infty }\protect \overline  \rho _{\protect \rm  red}$, see the discussion in the introduction about the commutativity of the two limits.}
\end{thebibliography}
\end{document}